\documentclass[12pt]{article}

\newcommand{\bs}{\boldsymbol}

\usepackage{amssymb,amsbsy,amsmath,amsthm}
\usepackage{tcolorbox}
\usepackage{epsf,epsfig}
\usepackage{texdraw}
\usepackage{bbding,pifont}

\newtheorem{theorem}{Theorem}

\newtheorem*{criterion*}{Criterion}
\newtheorem*{definition*}{Definition}
\newtheorem{definition}{Definition}

\theoremstyle{remark} 
\newtheorem{remark}{Remark}

\title{Systems of difference equations, symmetries and integrability conditions}
\author{Louis Brady\footnote{Email: 18005905@hope.ac.uk, Louis.Brady@outlook.com} and Pavlos Xenitidis\footnote{Email: xenitip@hope.ac.uk}\\ School of Mathematics, Computer Science \& Engineering\\ Liverpool Hope University, L16 9JD  Liverpool, UK}
\date{\today}

\begin{document}
	
	\maketitle
	\begin{abstract}
		We consider a class of systems of difference equations defined on an elementary quadrilateral of the ${\mathbb{Z}}^2$ lattice, define their eliminable and dynamical variables, and demonstrate their use. Using the existence of infinite hierarchies of symmetries as integrability criterion, we derive necessary integrability conditions and employ them in the construction of the lowest order symmetries of a given system. These considerations are demonstrated with the help of three systems from the class of systems under consideration. 
	\end{abstract}
	
	\section{Introduction}
	
The theory of symmetries of scalar difference equations is well developed and several methods and approaches have been proposed for their derivation, see for instance \cite{X1} and references therein. The existence of an infinite hierarchy of symmetries serves as an integrability criterion and it provides us with conservation laws via the so-called integrability conditions. The derivation of the latter relies on the existence of a formal recursion operator \cite{MWX,MX}, and some of these conditions can be interpreted as determining equations for the symmetries of the equation under consideration \cite{X1}.
	
The integrability of a system of difference equations is defined in the same way using the existence of infinite hierarchies of symmetries as a criterion. However, there is no systematic method to compute the symmetries of a given system therefore we have to employ other integrability aspects for their derivation. For instance there exist examples of integrable systems, the symmetries of which were derived using either the direct linearisation method \cite{TN} or a Lax pair \cite{FX}. The purpose of this paper is to contribute in the development of the theory of integrability conditions for systems of difference equations and the main aim is to propose a systematic method for the derivation of the lowest order symmetries of such systems using only the system under consideration.  
	
The systems of partial difference equations we are going to consider are defined on an elementary quadrilateral of the ${\mathbb{Z}}^2$ lattice and are determined by certain solvability requirements given in Definition \ref{def:class}. We call the variables involved in a system eliminable and dynamical when the former can be expressed uniquely in terms of the latter using the system. The importance of these variables stems from the fact that they play an important role in functional equations and the solution of the initial value problem, which is demonstrated with the use of three illustrative examples.
	
	We define formal Laurent and Taylor series of matrix pseudo-difference operators and discuss how they can be computed systematically. We employ them in the derivation of integrability conditions and explain how some of the derived conditions can be employed as determining equations for the symmetries of a given system. We demonstrate these interpretations using three systems, namely the coupled system of discrete potential Korteweg-de Vries (KdV) equations \eqref{eq:sys1}, the discrete nonlinear Schr\"odinger system \eqref{eq:sys2} \cite{KMX}, and the discrete Toda system \eqref{eq:sys3} \cite{A1}.
	
	The paper is organised as follows. The next section introduces our notation, defines the class of systems we are interested in along with eliminable and dynamical variables, and contains all the necessary definitions on symmetries. Section \ref{sec:recop} deals with recursion operators, formal series and integrability conditions, and discusses the implementation of our theory with the help of the three aforementioned systems. The concluding section contains an overall evaluation of our results and a discussion of further developments.

	\section{Systems of difference equations and symmetries}
	
	In this section we present our notation and define the class $\cal{Q}$ of systems of difference equations we are going to consider. We introduce the notion of eliminable and dynamical variables, and give the basic definitions of symmetries and conservation laws for difference equations. We present three systems from the class  $\cal{Q}$ to exemplify the concepts we introduce and use them as running examples throughout the paper.
	
	\subsection{Notation}
	
	In what follows we deal with systems of partial difference equations in two independent discrete variables denoted by $n$ and $m$. The dependence of a function $u$ on $n$ and $m$ is denoted with indices omitting the independent variables, i.e.,
	$$u(n+i,m+j) = u_{i,j},\qquad \forall ~ i,j \in {\mathbb{Z}}.$$
	The two shift operators related to the two lattice directions are 
	$${\cal{S}} : n \mapsto n+1,\quad {\cal{T}} : m \mapsto m+1,$$
	and their action on a function of $n$ and $m$ is defined as
	$${\cal{S}}^k {\cal{T}}^\ell (f_{i,j}) = f_{i+k,j+\ell} .$$
	Vectors are written in bold face letters and their components are distinguished by upper indices in parentheses, i.e.,
	$$ {\bs{f}}_{0,0} = \left( f^{(1)}_{0,0},\ldots,f^{(k)}_{0,0} \right). $$
	Matrices are denoted with upper case Roman letters. In particular, $\rm{I}$ and $\rm{O}$ denote the identity and the zero matrix, respectively. We also use the notation ${\bs{f}}([{\bs{u}}])$ to denote a vector function depending on a finite, but otherwise unspecified number of shifted values of $\bs{u}$, e.g., a vector function like $ {\bs{f}}({\bs{u}}_{-1,0},{\bs{u}}_{0,0},{\bs{u}}_{1,0})$. 
	
	\subsection{A class of systems of difference equations}
	
	We consider systems of $k$ equations for $k$ functions $u^{(i)} : {\mathbb{Z}}^2 \mapsto {\mathbb{R}}$, $i=1,\ldots,k$, which involve the values of these functions at four points which form an elementary quadrilateral on the $\mathbb{Z}^2$ lattice. That is systems of the form
	\begin{equation} \label{eq:sys}
		{\bs{Q}}({\bs{u}}_{0,0},{\bs{u}}_{1,0},{\bs{u}}_{0,1},{\bs{u}}_{1,1}) ={\bs{0}},
	\end{equation}
	where ${\bs{u}}_{i,j} = \left( u^{(1)}_{i,j},\ldots,  u^{(k)}_{i,j}\right)$.  We can also write system \eqref{eq:sys} in component form as
	$$Q^{(i)}({\bs{u}}_{0,0},{\bs{u}}_{1,0},{\bs{u}}_{0,1},{\bs{u}}_{1,1}) =0,\quad i =1,\ldots,k.$$
	We refer to system \eqref{eq:sys} as {\emph{a quad system}}. 
	
\begin{center}
		\begin{figure}[ht]
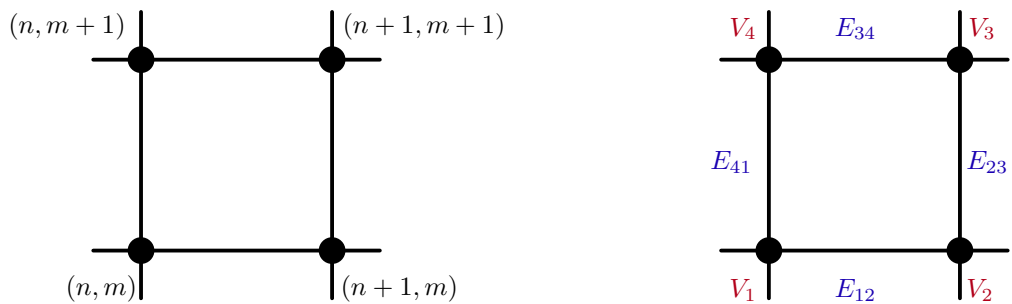
		
			\begin{minipage}{.5 \linewidth}
				\centertexdraw{ \setunitscale 1
					\linewd 0.02 \arrowheadtype t:F 
					\htext(0 0.5) {\phantom{T}}
					\move (-.25 0) \lvec(1.25 0)
					\move (0 -.25) \lvec (0 1.25) 
					\move (1 -.25) \lvec (1 1.25)
					\move (-.25 1) \lvec (1.25 1)
					\move (1 0) \fcir f:0.0 r:0.07 \move (0 0) \fcir f:0.0 r:0.07 \move (0 1) \fcir f:0.0 r:0.07
					\move (1 1) \fcir f:0.0 r:0.07 
					\htext (-.4 -.27) {\footnotesize{$(n,m)$}} \htext (1.04 -.27) {\footnotesize{$(n+1,m)$}}
					\htext (-.7 1.1) {\footnotesize{$(n,m+1)$}} \htext (1.05 1.1) {\footnotesize{$(n+1,m+1)$}}			
				}
			\end{minipage}
			\begin{minipage}{.5 \linewidth}
				\centertexdraw{ \setunitscale 1
					\linewd 0.02 \arrowheadtype t:F 
					\htext(0 0.5) {\phantom{T}}
					\move (-.25 0) \lvec(1.25 0)
					\move (0 -.25) \lvec (0 1.25) 
					\move (1 -.25) \lvec (1 1.25)
					\move (-.25 1) \lvec (1.25 1)
					\move (1 0) \fcir f:0.0 r:0.07 \move (0 0) \fcir f:0.0 r:0.07 \move (0 1) \fcir f:0.0 r:0.07
					\move (1 1) \fcir f:0.0 r:0.07 
					\htext (-.2 -.27) {{\color[rgb]{.7,0,.1}{\footnotesize{$V_1$}}}} \htext (1.04 -.27) {{\color[rgb]{.7,0,.1}{\footnotesize{$V_2$}}}}
					\htext (-.2 1.1) {{\color[rgb]{.7,0,.1}{\footnotesize{$V_4$}}}} \htext (1.05 1.1) {{\color[rgb]{.7,0,.1}{\footnotesize{$V_3$}}}}			
					\htext (.35 -.27) {{\color[rgb]{.1,0,.7}{\footnotesize{$E_{12}$}}}} \htext (1.04 .4) {{\color[rgb]{.1,0,.7}{\footnotesize{$E_{23}$}}}}
					\htext (-.3 0.4) {{\color[rgb]{.1,0,.7}{\footnotesize{$E_{41}$}}}} \htext (.35 1.1) {{\color[rgb]{.1,0,.7}{\footnotesize{$E_{34}$}}}}
				}
			\end{minipage}
			\caption{The elementary quadrilateral of the ${\mathbb{Z}}^2$ lattice where quad system \eqref{eq:sys} is defined and the naming of its vertices and edges.} \label{fig:quad}
		\end{figure}
	\end{center}

	Quad systems form a very wide class of systems, so we restrict our considerations to quad systems having certain properties which are described in the following definition. We denote the class of all these systems with $\cal{Q}$.
	\begin{definition}[Class of quad systems] \label{def:class}
		Let $V_i$, $i=1,\ldots,4$, denote the four vertices of an elementary quadrilateral of the $\mathbb{Z}^2$ lattice where quad system \eqref{eq:sys} is defined, and $E_{ij}$, or $E_{ji}$, denote the undirectional edge connecting two consecutive vertices $V_i$ and $V_j$ as shown in Figure \ref{fig:quad}. Let $\cal{U}$ denote the set of all variables appearing in quad system \eqref{eq:sys}.
		
		Class $\cal{Q}$ contains quad systems \eqref{eq:sys} which fulfill the following two requirements. 
		\begin{enumerate}
			\item[(i)] For every ${\bs{u}}_{p,q}$, $p,q=0,1$, the defining function $\bs{Q}$ of system \eqref{eq:sys} satisfies  
			\begin{equation} \label{eq:req}
				{\rm{Q}}_{(p,q)} \ne {\rm{O}} ~~~ \mbox{for all }~p,q=0,1,
			\end{equation}
			where ${\rm{Q}}_{(p,q)}$ are the Jacobian matrices of $\bs{Q}$. 
			\begin{equation}\label{eq:defJac}
				{\rm{Q}}_{(p,q)}  = \frac{\partial {\bs{Q}}}{\partial {\bs{u}}_{p,q}}  = \left(\frac{\partial {\bs{Q}}}{\partial u^{(1)}_{p,q}} \cdots \frac{\partial {\bs{Q}}}{\partial u^{(k)}_{p,q}}\right), \quad {\mbox{with }} \left({{\rm{Q}}_{(p,q)}}\right)_{a,b} = \frac{\partial Q^{(a)}}{\partial u^{(b)}_{p,q}}
			\end{equation} 
			\item[(ii)] For every edge $E_{ij}$ it can be solved uniquely for $\ell$ variables $(u^{(a_1)},\ldots,u^{(a_\ell)})$ lying on the vertex $V_i$ and $k-\ell$ variables $(u^{(b_1)},\ldots,u^{(b_{k-\ell})})$ lying on vertex $V_j$, where $0 \le \ell \le k$, and the resulting expressions involve all the variables in ${\cal{U}}- \{u^{(a_1)}_{V_i},\ldots,u^{(a_\ell)}_{V_i},u^{(b_1)}_{V_j},\ldots,u^{(b_{k-\ell})}_{V_j}\}$,  where $u^{(a)}_{V_s}$ denotes the value of the function $u^{(a)}$ at the vertex $V_s$.
		\end{enumerate}
	\end{definition}

	\begin{definition}
		We call variables $\{u^{(a_1)}_{V_i},\ldots,u^{(a_\ell)}_{V_i},u^{(b_1)}_{V_j},\ldots,u^{(b_{k-\ell})}_{V_j}\}$ {\emph{eliminable}} and all the other ones {\emph{dynamical}}.
	\end{definition}

	We call these variables eliminable because we can eliminate them from {\emph{functional relations and equations}}, like conservation laws and determining equations for symmetries, which must hold on solutions of  the quad system. They are also related to the initial value problem since the dynamical variables may be viewed as initial conditions and the eliminable ones as the values of the solution to be determined in terms of the initial data. These concepts and interpretations are  illustrated below with the help of three examples.
	
	\subsection{Symmetries and conservation laws}
	
	We define a symmetry of a system of difference equations in the following way.
	
	\begin{definition}
		A vector function 
		\begin{equation}\label{eq:gensym1}
			{\bs{F}}(n,m,[{\bs{u}}]) = \left( F^{(1)}(n,m,[{\bs{u}}]), \ldots,F^{(k)}(n,m,[{\bs{u}}])\right) 
		\end{equation}
		is {\emph{a symmetry generator}}, or just {\emph{a symmetry}}, of system ${\bs{Q}}(n,m,[{\bs{u}}])= {\bs{0}}$ if
		\begin{equation}\label{eq:gendeteq}
			\sum_{p,q} {\rm{Q}}_{(p,q)}  {\cal{S}}^p {\cal{T}}^q({\bs{F}}) = {\bs{0}} ~ {\mbox{ holds on solutions of }}~ {\bs{Q}}(n,m,[{\bs{u}}])= {\bs{0}},
		\end{equation}
		where matrices ${\rm{Q}}_{(p,q)}$ are defined in \eqref{eq:defJac}. Equations \eqref{eq:gendeteq} are referred to as the determining equations for the symmetries of ${\bs{Q}}(n,m,[{\bs{u}}])= {\bs{0}}$.
	\end{definition}
	If the system depends on a parameter $\alpha$ we can also consider symmetries which act on the parameter.
	\begin{definition}
		Let system ${\bs{Q}}(n,m,[{\bs{u}}];\alpha)= {\bs{0}}$ depend also on a parameter $\alpha$. Then, a vector function 
		\begin{equation}\label{eq:genmastsym1}
			\left({\bs{M}}(n,m,[{\bs{u}}]),\xi \right) = \left( M^{(1)}(n,m,[{\bs{u}}]), \ldots,M^{(k)}(n,m,[{\bs{u}}]),\xi \right)
		\end{equation}
		is an {\emph{extended symmetry generator}}, or just {\emph{an extended symmetry}}, of the system if
		\begin{equation}\label{eq:extgendeteq}
			\sum_{p,q} {\rm{Q}}_{(p,q)}  {\cal{S}}^p {\cal{T}}^q({\bs{M}}) + \xi \partial_\alpha {\bs{Q}} = {\bs{0}} ~ {\mbox{ holds on solutions of }}~ {\bs{Q}}(n,m,[{\bs{u}}];\alpha)= {\bs{0}}.
		\end{equation}
	\end{definition}
	With every symmetry we can associate a vector field. More precisely, with symmetry $\bs{F}$ \eqref{eq:gensym1} and extended symmetry $\bs{M}$ \eqref{eq:genmastsym1} we can associate the vector fields
	$$X_{\bs{F}} = \sum_{i=1}^{k} \sum_{p,q} {\cal{S}}^p{\cal{T}}^q\left(F^{(i)}\right) \frac{\partial {\phantom{u_{p,q}^{(i)}}}}{\partial u_{p,q}^{(i)}} \quad {\mbox{and}} \quad X_{\bs{M}} = \sum_{i=1}^{k} \sum_{p,q} {\cal{S}}^p{\cal{T}}^q\left(M^{(i)}\right) \frac{\partial {\phantom{u_{p,q}^{(i)}}}}{\partial u_{p,q}^{(i)}}  + \xi \frac{\partial {\phantom{\alpha}}}{\partial \alpha},$$
	respectively. The commutator $ [~,~]$ of two symmetries $\bs{F}$ and $\bs{G}$ results to a new symmetry $\bs{H}$ which is defined  as
	\begin{subequations} \label{eq:commutator}
		\begin{equation}
			{\bs{H}} = [{\bs{F}},{\bs{G}}] = X_{\bs{F}}({\bs{G}}) - X_{\bs{G}}({\bs{F}}),
		\end{equation}
		and its components are given by
		\begin{equation}
			H^{(i)} = \sum_{\ell_1,\ell_2} \sum_{j=1}^{k} {\cal{S}}^{\ell_1} {\cal{T}}^{\ell_2} \left( F^{(j)}\right) \frac{\partial G^{(i)}}{\partial u^{(j)}_{\ell_1,\ell_2}}  - {\cal{S}}^{\ell_1} {\cal{T}}^{\ell_2} \left(G^{(j)}\right) \frac{\partial F^{(i)}}{\partial u^{(j)}_{\ell_1,\ell_2}}.
		\end{equation} 
	\end{subequations}
	If ${\bs{H}} = {\bs{0}}$ then we say that the two symmetries commute.
	
	A generalised symmetry is called {\emph{master symmetry}} if its commutator with a generalised symmetry produces another generalised symmetry, and its existence may be used as an integrability criterion.
	
	\begin{remark}
		Using the relation of symmetries and continuous groups of transformations we can represent the former as systems of differential-difference equations. Assuming that $\bs{u}$ depends on two continuous variables $t$ and $\tau$ (the group parameters),  we can write symmetry generator \eqref{eq:gensym1} as 
		\begin{equation}\label{eq:gensym}
			\frac{\partial {\bs{u}}_{0,0}}{\partial t}  = {\bs{F}}(n,m,[{\bs{u}}]) \quad \Rightarrow \quad \frac{\partial u^{(i)}_{0,0}}{\partial t} = F^{(i)}(n,m,[{\bs{u}}]), ~~ i=1,\ldots,k,
		\end{equation}
		and extended symmetry generator \eqref{eq:genmastsym1} as
		\begin{equation}\label{eq:genmastsym}
			\frac{\partial {\bs{u}}_{0,0}}{\partial \tau}  = {\bs{M}}(n,m,[{\bs{u}}]), ~~ \frac{\partial \alpha}{\partial \tau}  = \xi \quad \Rightarrow \quad \frac{\partial u^{(i)}_{0,0}}{\partial \tau} = M^{(i)}(n,m,[{\bs{u}}]), ~~ i=1,\ldots,k, ~~ \frac{\partial \alpha}{\partial \tau}  = \xi.
		\end{equation}
		In what follows we denote symmetries as differential-difference equations.
	\end{remark}
	
	The sums in all the above definitions are finite. In particular, for quad systems \eqref{eq:sys} the determining equations \eqref{eq:gendeteq} involve only four terms,
	\begin{equation} \label{eq:deteq}
		{\rm{Q}}_{(0,0)} {\bs{F}} + {\rm{Q}}_{(1,0)} {\cal{S}}\left({\bs{F}}\right)+{\rm{Q}}_{(0,1)}{\cal{T}}\left( {\bs{F}}\right)+{\rm{Q}}_{(1,1)} {\cal{S T}} \left({\bs{F}}\right) = {\bs{0}},
	\end{equation}	
	and similarly \eqref{eq:extgendeteq} involves only five terms,
	\begin{equation} \label{eq:extdeteq}
		{\rm{Q}}_{(0,0)} {\bs{F}} + {\rm{Q}}_{(1,0)} {\cal{S}}\left({\bs{F}}\right)+{\rm{Q}}_{(0,1)}{\cal{T}}\left( {\bs{F}}\right)+{\rm{Q}}_{(1,1)} {\cal{S T}} \left({\bs{F}}\right) + \xi \partial_\alpha {\bs{Q}} = {\bs{0}}.
	\end{equation}	
	It can also be shown that the symmetries of \eqref{eq:sys} necessarily have the form  
	$$  {\bs{F}} = {\bs{A}}(n,m,{\bs{u}}_{-M,0},\ldots,{\bs{u}}_{M,0}) + {\bs{B}}(n,m,{\bs{u}}_{0,-M},\ldots,{\bs{u}}_{0,M}),$$
	where $M$ is a positive integer also known as {\emph{order of the symmetry}}. This allows us to consider symmetries involving shifts of $\bs{u}$ only in one direction. In what follows we discuss only {\emph{first order symmetries in the first direction}}, i.e. symmetries of the form ${\bs{F}} = {\bs{F}}(n,m,{\bs{u}}_{-1,0},{\bs{u}}_{0,0},{\bs{u}}_{1,0})$.

	\begin{definition*}
		A conservation law of system \eqref{eq:gendeteq} is a pair of functions $(\rho,\sigma) = (\rho(n,m,[{\bs{u}}]),\sigma(n,m,[{\bs{u}}]))$ such that $({\cal{T}}-1)(\rho) = ({\cal{S}}-1)(\sigma)$ on solutions of the system.  \\ A conservation law is trivial if $\rho = ({\cal{S}}-1)(H)$ and $\sigma = ({\cal{T}}-1)(H)$ for some function $H = H(n,m,[{\bs{u}}])$.
	\end{definition*}
	Conservation laws are related to symmetries and integrability as they may be derived from the integrability conditions as we explain in the next section.

	\subsection{Examples}
	
	In order to simplify our notation and avoid the use of many indices in the three running examples throughout this paper, we use the following notation for the unknown functions
	$$ {\bs{u}} = (u^{(1)},u^{(2)})  = (u,v).$$
	
	\subsubsection{First example}
	
	The first example is about the system
	\begin{equation}  \label{eq:sys1}
		(u_{0,0}  - u_{1,1})(v_{1,0} -v_{0,1})  - \alpha+ \beta = 0, \qquad  (v_{0,0}  - v_{1,1})(u_{1,0} -u_{0,1})  - \alpha+ \beta = 0.
	\end{equation}
	It involves all eight components of $ {\bs{u}}$ and its shifts, i.e. $ {\cal{U}} = \left\{u_{0,0},v_{0,0},u_{1,0},v_{1,0},u_{0,1},v_{0,1},u_{1,1},v_{1,1}\right\}$, and condition \eqref{eq:req} is readily satisfied because none of the Jacobian matrices are the zero matrix.
	\begin{equation} \label{eq:matQex1}
		{\rm{Q}}_{(0,0)}=- {\rm{Q}}_{(1,1)}= \begin{pmatrix}
			v_{1,0}-v_{0,1} & 0 \\
			0 & u_{1,0}-u_{0,1}
		\end{pmatrix},~ {\rm{Q}}_{(1,0)}=- {\rm{Q}}_{(1,0)} = \begin{pmatrix}
			0 & u_{0,0} - u_{1,1} \\ v_{0,0}-v_{1,1} & 0
		\end{pmatrix}.
	\end{equation}
	For the second requirement in Definition \ref{def:class}, we start by considering edge $E_{12}$, see Figure \ref{fig:quad}. It is clear that we can solve system \eqref{eq:sys1} uniquely for either $(u_{V_1},v_{V_1}) = (u_{0,0},v_{0,0})$ or  $(u_{V_2},v_{V_2}) = (u_{1,0},v_{1,0})$, and in both cases, it is not difficult to check that the resulting expressions involve all the other variables appearing in system \eqref{eq:sys1}.
	
	In the same way, for edge $E_{23}$ we can solve system \eqref{eq:sys1} uniquely for either $(u_{V_2},v_{V_2}) = (u_{1,0},v_{1,0})$ or  $(u_{V_3},v_{V_3}) = (u_{1,1},v_{1,1})$. Working similarly with the remaining two edges we can clearly conclude that system \eqref{eq:sys1} can be solved uniquely for any pair $(u_{i,j},v_{i,j})$, $i,j=0,1$, and all the requirements in Definition \ref{def:class} are fulfilled. Thus, for every edge we have two different choices for eliminable variables, and the set of all possible pairs is
	\begin{equation} \label{eq:elim1}
		{\cal{E}} = \{ (u_{0,0},v_{0,0}), ~ (u_{1,0},v_{1,0}),~ (u_{1,1},v_{1,1}), ~ (u_{0,1},v_{0,1})\}.
	\end{equation}
	This shows that system \eqref{eq:sys1} belongs in class $\cal{Q}$.
	
	Set $\cal{E}$ contains  all the possible pairs of variables which we can replace in a unique way from a given expression using system \eqref{eq:sys1} and it is useful in computations with symmetries and conservation laws. For instance, to show that the pair 
	\begin{equation}\label{eq:ex1cl}
		(\rho,\sigma) = (u_{0,0} v_{1,0} - u_{1,0} v_{0,0}\, , ~ u_{0,0} v_{0,1} - u_{0,1} v_{0,0})
	\end{equation}
	is a conservation law of system \eqref{eq:sys1}, it is sufficient to show that $({\cal{T}}-1)(\rho)-   ({\cal{S}}-1)(\sigma) = 0$, or in explicit form
	\begin{equation}\label{eq:ex1cl1}
		u_{0,1} v_{1,1} - u_{1,1} v_{0,1}- u_{0,0} v_{1,0} + u_{1,0} v_{0,0} -  u_{1,0} v_{1,1} + u_{1,1} v_{1,0}  +  u_{0,0} v_{0,1} - u_{0,1} v_{0,0} =0,
	\end{equation}
	modulo system \eqref{eq:sys1}. First we choose a pair from \eqref{eq:elim1} which we want to eliminate from \eqref{eq:ex1cl1}, e.g. $(u_{1,1},v_{1,1})$, and then rearrange system \eqref{eq:sys1} for these two variables.
	$$u_{1,1} = u_{0,0} - \frac{\alpha-\beta}{v_{1,0}-v_{0,1}},~~~ v_{1,1} = v_{0,0} - \frac{\alpha-\beta}{u_{1,0}-u_{0,1}}.$$
	Finally we use these relations to replace $u_{1,1}$, $v_{1,1}$ in \eqref{eq:ex1cl1} and turn the latter into an identity.

	\subsubsection{Second example}
	
	The second example is about a system derived in \cite{KMX} as a discretization of the nonlinear Schr\"odinger equation, namely
	\begin{equation} \label{eq:sys2}
		u_{1,0} - u_{0,1} - \frac{\alpha-\beta}{1+ u_{0,0} v_{1,1}} u_{0,0} = 0,\qquad v_{1,0} - v_{0,1} + \frac{\alpha-\beta}{1+ u_{0,0} v_{1,1}} v_{1,1} = 0.
	\end{equation}
	This system depends only on six of the eight components of $ {\bs{u}}$ and its shifts, as $u_{1,1}$ and $v_{0,0}$ do not appear in \eqref{eq:sys2}, hence $ {\cal{U}} = \left\{u_{0,0},u_{1,0},v_{1,0},u_{0,1},v_{0,1},v_{1,1}\right\}$. The Jacobian matrices 
	\begin{equation} \label{eq:matQex2}
		{\rm{Q}}_{(0,0)}= f_{0,0} \begin{pmatrix}
			1 & 0 \\
			v_{1,1}^2 & 0
		\end{pmatrix},\quad {\rm{Q}}_{(1,0)} = {\rm{I}}, \quad {\rm{Q}}_{(0,1)} = -{\rm{I}},\quad {\rm{Q}}_{(1,1)}=f_{0,0} \begin{pmatrix}
			0 & -u_{0,0}^2 \\ 0 & -1
		\end{pmatrix} ,
	\end{equation}
	where $ f_{0,0} = \frac{\beta-\alpha}{(1+ u_{0,0} v_{1,1})^2}$, are non-zero, thus condition \eqref{eq:req} is satisfied. 
	
	It is clear that this system can be solved uniquely for  the vertices $(u_{V_2},v_{V_2}) = (u_{1,0},v_{1,0})$ and $(u_{V_4},v_{V_4}) = (u_{0,1},v_{0,1})$. However there are more choices for eliminable variables. For edge $E_{12}$, we can solve system \eqref{eq:sys2} uniquely also for $(u_{V_1},v_{V_2}) = (u_{0,0},v_{1,0})$. For edge $E_{23}$, we have another pair, namely $(u_{V_2},v_{V_3}) = (u_{1,0},v_{1,1})$. Finally, for edges $E_{34}$ and  $E_{41}$  we find $(u_{V_4},v_{V_3}) = (u_{0,1},v_{1,1})$ and $(u_{V_1},v_{V_4}) = (u_{0,0},v_{0,1})$ respectively. Summarising, for every edge we have two different choices of eliminable variables, and for every choice the second requirement in Definition \ref{def:class} is fully satisfied. Thus the system belongs in class $\cal{Q}$ and the set of all possible eliminable variables is
	\begin{equation} \label{eq:elim2}
		{\cal{E}} = \{ (u_{0,0},v_{1,0}), ~ (u_{1,0},v_{1,0}), ~ (u_{1,0},v_{1,1}), ~ (u_{0,1},v_{1,1}), ~ (u_{0,1},v_{0,1}),~ (u_{0,0},v_{0,1}) \}.
	\end{equation}

\begin{center}
		\begin{figure}[th]
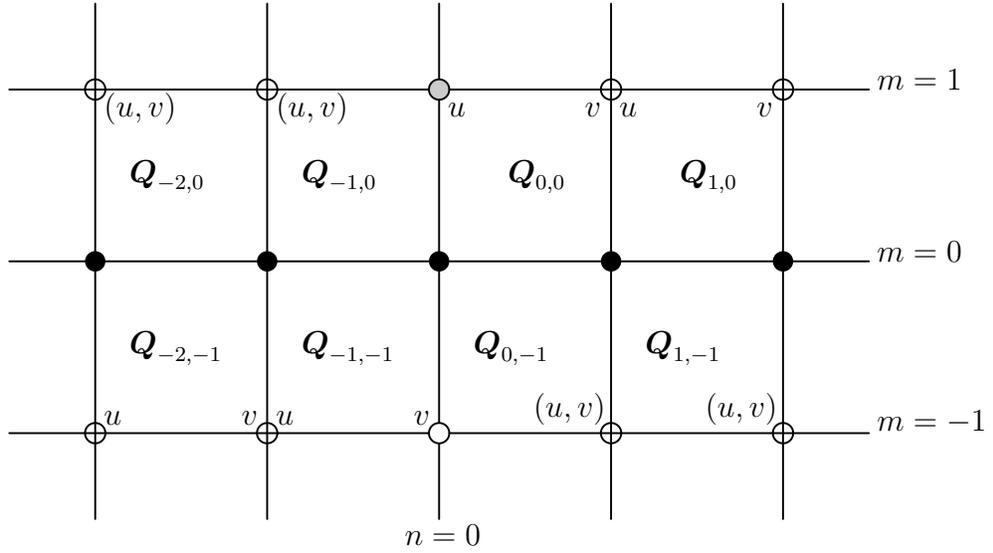

			\centertexdraw{ \setunitscale .9 \linewd 0.01 \arrowheadtype t:F
				\move(-2.5 0)  \lvec(2.5 0) 
				\move(-2.5 1)  \lvec(2.5 1)
				\move(-2.5 -1)  \lvec(2.5 -1)  
				\move(1 -1.5)  \lvec(1 1.5) 			
				\move(-1 -1.5)  \lvec(-1 1.5) 			
				\move(2 -1.5)  \lvec(2 1.5) 			
				\move(-2 -1.5)  \lvec(-2 1.5) 			
				\move(0 -1.5)  \lvec(0 1.5) 	
				
				\move(0 0) \fcir f:0 r:0.06
				\move(1 0) \fcir f:0 r:0.06
				\move(-1 0) \fcir f:0 r:0.06
				\move(2 0) \fcir f:0 r:0.06
				\move(-2 0) \fcir f:0 r:0.06
				
				\move(0 1) \fcir f:0.8 r:0.06
				\move(0 1) \lcir r:0.06
				
				\move(0 -1) \fcir f:1 r:0.06
				\move(0 -1) \lcir r:0.06
				
				\move(1 1) \lcir r:0.06
				\move(1 -1) \lcir r:0.06
				\move(-1 1) \lcir r:0.06
				\move(-1 -1) \lcir r:0.06
				\move(2 1) \lcir r:0.06
				\move(2 -1) \lcir r:0.06
				\move(-2 1) \lcir r:0.06
				\move(-2 -1) \lcir r:0.06
				
				\htext (0.4 .4) {${\bs{Q}}_{0,0}$} 
				\htext (-.8 .4) {${\bs{Q}}_{-1,0}$} 
				\htext (0.2 -.6) {${\bs{Q}}_{0,-1}$}
				\htext (-.8 -.6) {${\bs{Q}}_{-1,-1}$}
				\htext (1.4 .4) {${\bs{Q}}_{1,0}$} 
				\htext (-1.8 .4) {${\bs{Q}}_{-2,0}$} 
				\htext (1.2 -.6) {${\bs{Q}}_{1,-1}$}
				\htext (-1.8 -.6) {${\bs{Q}}_{-2,-1}$}
				
				\htext (2.55 0) {$m=0$}  \htext (2.55 -1) {$m=-1$} \htext (2.55 1) {$m=1$}
				\htext (-0.20 -1.65) {$n=0$}
				\htext (0.05 .85) {$u$}  \htext (0.85 .85) {$v$} 
				\htext (1.05 .85) {$u$}  \htext (1.85 .85) {$v$} 
				\htext (-.95 .80) {$(u,v)$}	\htext (.55 -.95) {$(u,v)$}
				\htext (-1.95 .80) {$(u,v)$}	\htext (1.55 -.95) {$(u,v)$}
				\htext (-.95 -.95) {$u$}	\htext (-.15 -.95) {$v$}		 
				\htext (-1.95 -.95) {$u$}	\htext (-1.15 -.95) {$v$}		 
			}
			\caption{The initial value problem: initial values (dynamical variables) are denoted with disks ($u_{\pm i,0}$, $v_{\pm i,0}$ (black disks),  $v_{0,i}$ (grey disk), $u_{0,-i}$ (white disk), $i \in {\mathbb{Z}}_{\ge 0}$), and in every quadrilateral the eliminable variables $u$, $v$ which can be determined uniquely in terms of the initial data are given next to the corresponding vertex indicated with a circle.} \label{fig:sys2}
		\end{figure}
	\end{center}

	The dynamical variables may be viewed as initial conditions and the eliminable variables as the values of the solution to be determined in terms of the initial data. For system \eqref{eq:sys2} the initial value problem along a staircase was discussed in \cite{KMX} which actually only exploits the eliminable variables $(u_{1,0},v_{1,0})$ and $(u_{0,1},v_{0,1})$. However we can consider other initial value problems which exploit more than one set of eliminable variables. In order to simplify our presentation, let indices in $g_{i,j}$ denote the value of function $g$ at the lattice point $(i,j)$, i.e., $g_{i,j} = g(i,j)$. Suppose that initial values $\left\{ u(\pm i,0), ~v(\pm i,0), ~v(0,i), ~u(0,-i)\,;  ~ i \in {\mathbb{Z}}_{\ge 0}\right\}$ are given, see Figure \ref{fig:sys2}. Then we can determine the solution everywhere on the lattice using the system and its shifts. More precisely, using system ${\bs{Q}}_{0,0}=0$ we uniquely determine $(u_{0,1},v_{1,1})$ in terms of the initial data $u_{0,0}$, $u_{1,0}$, $v_{1,0}$ (black circles) and $v_{0,1}$ (grey circle). We continue by using systems ${\bs{Q}}_{j,0}=0$, $j = 1,2,\ldots$, and determine successively the corresponding pairs $(u_{j,1},v_{j+1,1})$. On the other hand, using systems ${\bs{Q}}_{-j,0}=0$, $j = 1,2,\ldots$,  we successively determine the pairs $(u_{-j,1},v_{-j,1})$. In this way we find the solution at $m=1$. We work similarly for negative values of $m$. Systems ${\bs{Q}}_{k,-1}= 0$, $k=0,1,\ldots$, yield successively the values  $(u_{k+1,-1},v_{k+1,-1})$, whereas systems ${\bs{Q}}_{-k,-1}= 0$, $k=1,2,\ldots$, yield values $(u_{-k,-1},v_{-k+1,-1})$. In this way we find the solution at $m=-1$.  Obviously we can repeat this procedure and determine the solution everywhere on the ${\mathbb{Z}}^2$ lattice.  
	
	\subsubsection{Third example}
	
	The last example is about the following quad system
	\begin{equation} \label{eq:sys3}
		u_{0,0} - v _{1,1} = 0,\qquad  (u_{1,0}-u_{0,0})(v_{1,0}-u_{0,0}) - (u_{0,1}-u_{0,0})(v_{0,1}- u_{0,0}) = 0,
	\end{equation}
	which upon the elimination of variable $v$ is equivalent to a discrete Toda system \cite{A1}. It does not depend on $u_{1,1}$, $v_{0,0}$, i.e. $ {\cal{U}} = \left\{u_{0,0},u_{1,0},v_{1,0},u_{0,1},v_{0,1},v_{1,1}\right\}$, and all of its Jacobian matrices are different from the zero matrix.
	\begin{subequations} \label{eq:matQex3}
		\begin{eqnarray}
			&& {\rm{Q}}_{(0,0)}= \begin{pmatrix}
				1 & 0 \\
				f_{0,0} & 0
			\end{pmatrix},\quad {\rm{Q}}_{(1,0)} = \begin{pmatrix} 0 & 0 \\ v_{1,0}-u_{0,0} & u_{1,0}-u_{0,0} \end{pmatrix}, \\ 
			&& {\rm{Q}}_{(0,1)} = \begin{pmatrix} 0 & 0 \\ u_{0,0}-v_{0,1} & u_{0,0}-u_{0,1} \end{pmatrix},\quad {\rm{Q}}_{(1,1)}= \begin{pmatrix}
				0 & -1 \\ 0 & 0
			\end{pmatrix} ,
		\end{eqnarray}
	\end{subequations}
	where $ f_{0,0} = u_{0,1}+v_{0,1} - u_{1,0}-v_{1,0}$. 
	
	It is not difficult to see that there is only one pair of eliminable variables for every edge, namely 
	\begin{equation} \label{eq:elim3}
		{\cal{E}} = \{(u_{0,0}, v_{1,0}), ~~ (u_{1,0},v_{1,1}), ~~ (u_{0,1},v_{1,1}), ~~ (u_{0,0},v_{0,1})\},
	\end{equation}
	and then check that the conditions of the second requirement in Definition \ref{def:class} are satisfied. Thus \eqref{eq:sys3} belongs in class $\cal{Q}$. 
	
\begin{center}
		\begin{figure}[th]
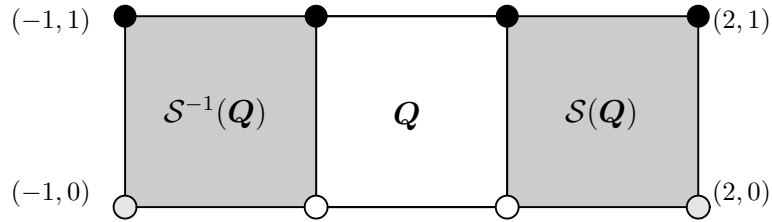

			\centertexdraw{ \setunitscale 1 \linewd 0.01 \arrowheadtype t:F
				\move(1 0)  \lvec(2 0) \lvec(2 1) \lvec(1 1) \lvec(1 0)
				\move(0 0)  \lvec(1 0) \lvec(1 1) \lvec(0 1) \lvec(0 0) \ifill f:0.8
				\move(2 0)  \lvec(3 0) \lvec(3 1) \lvec(2 1) \lvec(2 0) \ifill f:0.8
				\move(0 0) \lvec (2 0) \lvec (3 0)
				\move(0 1) \lvec (0 0) 
				\move(1 1) \lvec (1 0)
				\move(2 1) \lvec (2 0) 
				\move(3 1) \lvec (3 0)  
				\move(0 1) \lvec (2 1) \lvec (3 1)
				
				\move(1 1) \fcir f:0 r:0.06
				\move(2 1) \fcir f:0 r:0.06
				
				\move(0 0) \fcir f:0.9 r:0.06
				\move(0 0) \lcir r:0.06
				
				\move(1 0) \fcir f:1 r:0.06
				\move(1 0) \lcir r:0.06
				
				\move(0 1) \fcir f:0 r:0.06
				\move(3 1) \fcir f:0 r:0.06
				
				\move(2 0) \fcir f:1 r:0.06
				\move(2 0) \lcir r:0.06
				
				\move(3 0) \fcir f:0.9 r:0.06
				\move(3 0) \lcir  r:0.06
				
				\htext(-.6 0) {\footnotesize{$(-1,0)$}}
				\htext(-.6 .9) {\footnotesize{$(-1,1)$}}
				\htext(3.07 0) {\footnotesize{$(2,0)$}}
				\htext(3.07 .9) {\footnotesize{$(2,1)$}}
				\htext (1.4 .4) {${\bs{Q}}$} 	
				\htext (2.3 .4) {${\cal{S}}({\bs{Q}})$} 
				\htext (0.2 .4) {${\cal{S}}^{-1}({\bs{Q}})$} 
			}
			\caption{The three copies of the system, the eliminable variables (white and grey circles) and the dynamical ones (black circles).} \label{fig:sys3}
		\end{figure}
	\end{center}

	To prove that 
	$$ G^{(1)}_{0,0}  = (u_{1,0}-u_{0,0})(v_{1,0}-u_{0,0}), ~~~G^{(2)}_{0,0} = (u_{-1,0}-v_{0,0}) (v_{-1,0}-v_{0,0})$$
	is a symmetry of system \eqref{eq:sys3}, we have to show that the determining equations 
	\begin{eqnarray*}
		&& G^{(1)}_{0,0} - G^{(2)}_{1,1} =0, \\ 
		&& f_{0,0} G^{(1)}_{0,0} + (v_{1,0}-u_{0,0}) G^{(1)}_{1,0} + (u_{1,0}-u_{0,0}) G^{(2)}_{1,0} +  (u_{0,0}-v_{0,1}) G^{(1)}_{0,1} + (u_{0,0}-u_{0,1}) G^{(2)}_{0,1} = 0,
	\end{eqnarray*}
	hold on solutions of \eqref{eq:sys3}.  These equations involve 12 variables, namely $u_{-1,1}$, $v_{-1,1}$, $u_{0,0}$, $u_{1,0}$, $u_{0,1}$, $u_{1,1}$, $v_{0,0}$, $v_{0,1}$, $ v_{1,0}$, $v_{1,1}$, $ u_{2,0}$ and $v_{2,0}$, which are related via the system ${\bs{Q}}={\bs{0}}$ and its shifts  ${\cal{S}}({\bs{Q}})={\bs{0}}$,  ${\cal{S}}^{-1}({\bs{Q}})={\bs{0}}$, see Figure \ref{fig:sys3}. Using these three systems we can eliminate some of the variables in a consistent way. First we choose a pair of eliminable variables in \eqref{eq:elim3}, for instance $(u_{0,0},v_{1,0})$, which implies that we have also to eliminate its shifts $(u_{-1,0},v_{0,0})$ and $(u_{1,0},v_{2,0})$. Then we eliminate $(u_{-1,0},v_{0,0})$ using ${\cal{S}}^{-1}({\bs{Q}}) = {\bs{0}}$, next $(u_{0,0},v_{1,0})$ using ${\bs{Q}} = {\bs{0}}$, and finally $(u_{1,0}, v_{2,0})$ with the use of ${\cal{S}}({\bs{Q}}) = {\bs{0}}$. Upon the elimination of these variables, the determining equations become identities.
	
	\section{Recursion operators and integrability conditions} \label{sec:recop}
	
	The existence of an infinite hierarchy of generalised symmetries of increasing order serves as an integrability criterion. It has been successfully used in the study, analysis, and classification of differential, differential--difference and difference equations, see for instance \cite{LY,M,MWX,MX,S,X1,Y2}.  For the existence of a hierarchy in the $n$ direction it is sufficient to show that the system admits  {\emph{a recursion operator}} $\mathfrak{R}$ which maps symmetries to symmetries, i.e. an $\cal{S}$-pseudo-difference operator like
	$$ {\mathfrak{R}} = \begin{pmatrix} {\cal{S}} - 2 u_{0,0} v_{0,0} & -u_{0,0}^2 \\ v_{0,0}^2 & {\cal{S}}^{-1} \end{pmatrix} + 2 \begin{pmatrix} -u_{0,0} \\ v_{0,0} \end{pmatrix}({\cal{S}}-1)^{-1} \begin{pmatrix} v_{0,0} & u_{0,0} \end{pmatrix}.$$
	For more details about difference operators one may refer to \cite{MWX}. For the existence of a recursion operator for system \eqref{eq:sys} we state the following theorem without a proof because it is similar to the scalar case \cite{MWX}.
	\begin{theorem}
		An $\cal{S}$-pseudo-difference operator $\mathfrak{R}$ is a recursion operator for the symmetries in the $n$ direction of quad system \eqref{eq:sys} if 
		\begin{subequations} \label{eq:recop1}
			\begin{equation}\label{eq:recop1a}
				{\cal{T}}(\mathfrak{R})  ={\cal{B}}^{-1}\circ {\cal{A}} \circ {\mathfrak{R}}\circ {\cal{A}}^{-1} \circ {\cal{B}},
			\end{equation}
			where the linear operators $\cal{A}$ and $\cal{B}$ are given  by
			\begin{equation} \label{eq:defAB}
				{\cal{A}} = {\rm{Q}}_{(0,0)} + {\rm{Q}}_{(1,0)} {\cal{S}} ,\qquad {\cal{B}} = {\rm{Q}}_{(0,1)} + {\rm{Q}}_{(1,1)} {\cal{S}}.
			\end{equation}
		\end{subequations}
	\end{theorem}
	Operators $\cal{A}$ and $\cal{B}$ depend on the variables appearing in system \eqref{eq:sys}, whereas recursion operator $\mathfrak{R}$ depends only on variables ${\bs{u}}_{i,0}$. Moreover, ${\cal{T}}(\mathfrak{R})$ is not the composition of operators but the action of $\cal{T}$ on the functions appearing in the recursion operator, i.e. ${\cal{T}}(\mathfrak{R})$ is an $\cal{S}$-pseudo-difference operator depending  on variables ${\bs{u}}_{i,1}$. Thus relation \eqref{eq:recop1a} holds on solutions of the quad system under consideration. 
	
	An $\cal{S}$-pseudo-difference operator can be represented by a formal series, i.e. an infinite series in the shift operator $\cal{S}$. More precisely, we can represent an $\cal{S}$-pseudo-difference operator either with its {\emph{formal Laurent series}} which involves an infinite number of negative powers of $\cal{S}$ and maybe a finite number of positive powers, e.g.
	$$({\cal{S}}- {\rm{I}})^{-1} = {\cal{S}}^{-1} + {\cal{S}}^{-2} + {\cal{S}}^{-3} + \cdots, $$   
	or its {\emph{formal Taylor series}} which involves an infinite number of positive powers of $\cal{S}$ and maybe a finite number of negative powers, e.g.
	$$({\cal{S}}-{\rm{I}})^{-1} = -{\rm{I}} - {\cal{S}} - {\cal{S}}^{2} - {\cal{S}}^{3} - \cdots. $$
	The highest power of $\cal{S}$ in the Laurent series and the lowest power of $\cal{S}$ in the Taylor series of an operator are called the {\emph{Laurent and Taylor order}} of the operator, respectively.
	
	We can represent all the operators involved in \eqref{eq:recop1} with their formal series and derive equations which involve the coefficients of the formal series of the recursion operator.  Specifically, using indices $L$ (Laurent) and $T$ (Taylor) to indicate the formal series we are using, equation \eqref{eq:recop1} becomes
	\begin{subequations} \label{eq:recop2}
		\begin{eqnarray}
			&& {\cal{T}}({\mathfrak{R}}_L)  ={\cal{B}}^{-1}_L\circ {\cal{A}} \circ {\mathfrak{R}}_L\circ {\cal{A}}^{-1}_L \circ {\cal{B}}, \quad {\mbox{with }}~~{\mathfrak{R}}_L =\tilde{{\rm{R}}}_1 {\cal{S}} + \tilde{{\rm{R}}}_0  +  \tilde{{\rm{R}}}_{-1} {\cal{S}}^{-1} +  \ldots,  \label{eq:recop2a}\\ 
			&& {\cal{T}}({\mathfrak{R}}_T)  ={\cal{B}}^{-1}_T\circ {\cal{A}} \circ {\mathfrak{R}}_T\circ {\cal{A}}^{-1}_T \circ {\cal{B}}, \quad {\mbox{with }}~~{\mathfrak{R}}_T = \hat{{\rm{R}}}_{-1}  {\cal{S}}^{-1} + \hat{{\rm{R}}}_0  + \hat{{\rm{R}}}_1 {\cal{S}} +   \ldots . \label{eq:recop2b}
		\end{eqnarray}
	\end{subequations}
	The order of the recursion operator is related to the lowest order generalised symmetry admitted by the quad system. Since the coefficients of the formal series of $\mathfrak{R}$ are the unknowns in \eqref{eq:recop2} and we do not know any of the symmetries of the system, we assume that it is a first order operator which implies that the Laurent order of ${\mathfrak{R}}$ is one and its Taylor order is $-1$ as in \eqref{eq:recop2}.
	
	We have also to compute the formal inverse of operators $\cal{A}$ and $\cal{B}$ but this can be done systematically for a given quad system. Indeed, for a local operator ${\cal{P}} = {\rm{X}} + {\rm{Y}} {\cal{S}}$ with $\det({\rm{XY}}) \ne 0$, it is simple to derive the formal series of its inverse. If we assume that ${\cal{P}}^{-1}_L = {\rm{H}}_1 {\cal{S}}^{-1} + {\rm{H}}_2 {\cal{S}}^{-2} + \ldots$, then the requirement ${\cal{P}}^{-1}_L \circ {\cal{P}}  = {\rm{I}}$ leads to  ${\rm{H}}_1 {\cal{S}}^{-1}({\rm{Y}}) = {\rm{I}}$, ${\rm{H}}_i {\cal{S}}^{-i}({\rm{X}}) + {\rm{H}}_{i+1}{\cal{S}}^{-i-1}({\rm{Y}}) = {\rm{O}}$, for $i=1,2,\ldots$. Similarly for the Taylor series we start with ${\cal{P}}^{-1}_T = {\rm{Z}}_0 + {\rm{Z}}_1 {\cal{S}}+ {\rm{Z}}_2 {\cal{S}}^{2} + \ldots$ and the requirement ${\cal{P}}^{-1}_T \circ {\cal{P}} = {\rm{I}}$ leads to ${\rm{Z}}_0  {\rm{X = I}}$,  ${\rm{Z}}_i {\cal{S}}^{i}({\rm{Y}}) +  {\rm{Z}}_{i+1} {\cal{S}}^{i+1}({\rm{X}}) = {\rm{O}}$, $i=0,1,\ldots$.
	
	We can apply these considerations to compute the formal Laurent and Taylor inverse of operators ${\cal{A}} = {\rm{Q}}_{(0,0)} + {\rm{Q}}_{(1,0)} {\cal{S}}$ and $ {\cal{B}} =  {\rm{Q}}_{(0,1)} + {\rm{Q}}_{(1,1)} {\cal{S}}$ assuming that all the Jacobian matrices are invertible. Specifically, the formal Laurent series of ${\cal{A}}^{-1}$ and ${\cal{B}}^{-1}$ can be written as 
	\begin{subequations}\label{eq:invL}
		\begin{eqnarray}
			{\cal{A}}^{-1}_L &=&  {\cal{S}}^{-1}({\rm{Q}}_{(1,0)} ^{-1})  {\cal{S}}^{-1}  -  {\cal{S}}^{-1}\left({\rm{Q}}_{(1,0)} ^{-1} {\rm{Q}}_{(0,0)}   {\cal{S}}^{-1}({\rm{Q}}_{(1,0)} ^{-1})\right){\cal{S}}^{-2} + \ldots  , \label{eq:AinvL}\\
			{\cal{B}}^{-1}_L &=&  {\cal{S}}^{-1}({\rm{Q}}_{(1,1)} ^{-1})  {\cal{S}}^{-1}  -  {\cal{S}}^{-1}\left({\rm{Q}}_{(1,1)} ^{-1} {\rm{Q}}_{(0,1)}   {\cal{S}}^{-1}({\rm{Q}}_{(1,1)} ^{-1})\right){\cal{S}}^{-2} + \ldots , \label{eq:BinvL}
		\end{eqnarray}
	\end{subequations}
	and their formal Taylor series as
	\begin{subequations}\label{eq:invT}
		\begin{eqnarray}
			{\cal{A}}^{-1}_T &=&  {\rm{Q}}_{(0,0)} ^{-1}  -  {\rm{Q}}_{(0,0)} ^{-1} {\rm{Q}}_{(1,0)}   {\cal{S}}({\rm{Q}}_{(0,0)} ^{-1}){\cal{S}} + \ldots , \label{eq:AinvT} \\~~~
			{\cal{B}}^{-1}_T &=&  {\rm{Q}}_{(0,1)} ^{-1}  -  {\rm{Q}}_{(0,1)} ^{-1} {\rm{Q}}_{(1,1)}   {\cal{S}}({\rm{Q}}_{(0,1)} ^{-1}){\cal{S}} + \cdots . \label{eq:BinvT}
		\end{eqnarray}
	\end{subequations}
	
	We can work in the same way when $\det({\rm{XY}}) = 0$, but we have to use different ansatz for the formal series. For instance, if ${\rm{Y}}$ is singular then equation ${\rm{H}}_1 {\cal{S}}^{-1}({\rm{Y}}) = {\rm{I}}$ does not have solution, therefore we have to assume that ${\cal{P}}^{-1}_L = {\rm{H}}_0+ {\rm{H}}_1 {\cal{S}}^{-1} + {\rm{H}}_2 {\cal{S}}^{-2} + \ldots$ 	which leads to  ${\rm{H}}_0 {\rm{Y}} =  {\rm{O}}$, ${\rm{H}}_0 {\rm{X}} + {\rm{H}}_1 {\cal{S}}^{-1} ({\rm{Y}}) =  {\rm{I}}$, and  ${\rm{H}}_i {\cal{S}}^{-i} ({\rm{X}}) + {\rm{H}}_{i+1} {\cal{S}}^{-i-1} ({\rm{Y}}) =  {\rm{O}}$, for  $i =1,2,\ldots$. In the same way, if $\det({\rm{X}}) = 0$ then ${\rm{Z}}_0  {\rm{X = I}}$ does not have solution which implies that the formal Taylor series should be of the form  $	{\cal{P}}^{-1}_T = {\rm{Z}}_{-1} {\cal{S}}^{-1} + {\rm{Z}}_0 + {\rm{Z}}_1 {\cal{S}}+ {\rm{Z}}_2 {\cal{S}}^{2} + \ldots$ leading to ${\rm{Z}}_{-1} {\cal{S}}^{-1}({\rm{X}})=  {\rm{O}}$,  $ {\rm{Z}}_{-1} {\cal{S}}^{-1} ({\rm{Y}})+  {\rm{Z}}_0 {\rm{X}}  =  {\rm{I}}$,  and ${\rm{Z}}_i {\cal{S}}^{i} ({\rm{Y}}) + {\rm{Z}}_{i+1} {\cal{S}}^{i+1} ({\rm{X}}) =  {\rm{O}}$, for $i =0,1,\ldots$. But we can avoid the use of these different choices by rearranging equations \eqref{eq:recop2} appropriately as we demonstrate below in Subsection \ref{sec:singular}.
	
	The only other case we have to consider is when both matrices $\rm{X}$ and $\rm{Y}$ are singular, and the inverse may be a local operator 
	\begin{equation}\label{eq:invall}
		{\cal{P}}^{-1} =   {\rm{H}} {\cal{S}}^{-1} +  {\rm{Z}}, ~~~~{\rm{H}} {\cal{S}}^{-1}({\rm{X}}) = {\rm{O}}, ~~ {\rm{H}} {\cal{S}}^{-1}({\rm{Y}}) +  {\rm{Z}} {\rm{X}} = {\rm{I}}, ~~  {\rm{Z}} {\rm{Y}}= {\rm{O}}.
	\end{equation}
	
	In view of these substitutions, equations \eqref{eq:recop2} lead to (infinite) sets of determining relations for the coefficients of ${\mathfrak{R}}_L$ and  ${\mathfrak{R}}_T$, respectively. These relations are the {\emph{integrability conditions}} for quad system \eqref{eq:sys}. The reason we employ both series in our derivations is that they lead to integrability conditions which involve different derivatives of symmetry $\bs{F}$. Specifically, it can be shown \cite{X1} that the leading term $\tilde{{\rm{R}}}_1$ in the Laurent series is proportional to $\partial {\bs{F}}/\partial {\bs{u}}_{1,0}$ and we can consider that
	\begin{equation} \label{eq:RLsym}
		{\mathfrak{R}}_L = \frac{\partial {\bs{F}}}{\partial {\bs{u}}_{1,0}} {\cal{S}} + \tilde{{\rm{R}}}_0  +   \tilde{{\rm{R}}}_{1} {\cal{S}}^{-1} + \tilde{{\rm{R}}}_{2} {\cal{S}}^{-2} +\cdots, 
	\end{equation}
	whereas the leading term $\hat{{\rm{R}}}_{-1}$ in the Taylor series is proportional to $\partial {\bs{F}}/\partial {\bs{u}}_{-1,0}$ and thus
	\begin{equation} \label{eq:RTsym}
		{\mathfrak{R}}_T = \frac{\partial {\bs{F}}}{\partial {\bs{u}}_{-1,0}} {\cal{S}}^{-1} + \hat{{\rm{R}}}_0  +   \hat{{\rm{R}}}_{1} {\cal{S}} +   \hat{{\rm{R}}}_{2} {\cal{S}}^2+ \cdots.
	\end{equation}
	This connection between symmetries and formal recursion operators allows us to interpret the first integrability conditions also as {\emph{determining equations}} for the symmetry ${\bs{F}}$ along with \eqref{eq:deteq}. 
	
	\begin{remark} \label{rem:Rell}
		Using a first order symmetry along with \eqref{eq:RLsym} and \eqref{eq:RTsym}, we may consider equations \eqref{eq:recop2} with ${\mathfrak{R}}_L$ and ${\mathfrak{R}}_T$ being replaced by ${\mathfrak{R}}_L^\ell$ and ${\mathfrak{R}}_T^\ell$, respectively, where $\ell>1$ is an integer. The analysis of these equations leads to integrability conditions which can be used to determine the $\ell^{\rm{th}}$ order symmetries of the system. This approach is quite useful if we want to construct higher order symmetries and to prove integrability in the case when a master symmetry does not exist or the recursion operator cannot be computed in a closed form, see Section \ref{sec:3exsym} below.
	\end{remark}
	
	\subsection{First case: invertible Jacobian matrices}
	
	We begin the analysis of equations \eqref{eq:recop2} by considering systems for which all the Jacobian matrices ${\rm{Q}}_{(i,j)}$ are invertible. Such systems constitute a generalisation of the class of scalar quad equations which is very well studied, see for instance \cite{HJN} and references therein. In the same way, our analysis below may also be seen as a generalisation of the corresponding theory developed in \cite{MWX, MX}.

	Starting with \eqref{eq:defAB} and the Laurent series \eqref{eq:invL}, we can easily compute ${\cal{A}}^{-1}_L \circ {\cal{B}}$ and ${\cal{B}}^{-1}_L \circ {\cal{A}}$. If we write these operators as
	$${\cal{A}}^{-1}_L \circ {\cal{B}} = {\rm{B}}_0 + {\rm{B}}_1 {\cal{S}}^{-1} + {\rm{B}}_2 {\cal{S}}^{-2} + \ldots, \qquad  {\cal{B}}^{-1}_L \circ {\cal{A}} = {\rm{A}}_0 + {\rm{A}}_1 {\cal{S}}^{-1} + {\rm{A}}_2 {\cal{S}}^{-2} + \ldots, $$
	then equation \eqref{eq:recop2a} becomes
	$${\cal{T}}\left( \tilde{\rm{R}}_1 \right) {\cal{S}} + {\cal{T}}\left(\tilde{{\rm{R}}}_0 \right) +   \ldots = {\rm{B}}_0^{-1}\tilde{{\rm{R}}}_1 {\cal{S}}({\rm{B}}_0) {\cal{S}} + \left( {\rm{B}}_0^{-1} \tilde{{\rm{R}}}_1 {\cal{S}}({\rm{B}}_1)  +  {\rm{B}}_0^{-1} \tilde{{\rm{R}}} _0 {\rm{B}}_0 + {\rm{A}}_1 {\cal{S}}^{-1}( \tilde{{\rm{R}}}_1 ) {\rm{B}}_0 \right) + \ldots,$$
	where we have taken into account that ${\rm{A_0B_0 = B_0A_0 = I}}$.
	By equating coefficients of different powers of the shift operator $\cal{S}$ on both sides of this relation, we end up with an infinite set of integrability conditions, the first one of which is 
	\begin{equation} \label{eq:deteq1L}
		{\cal{T}}\left( \tilde{\rm{R}}_1 \right) =  {\rm{B}}_0^{-1}\tilde{{\rm{R}}}_1 {\cal{S}}({\rm{B}}_0), ~~{\mbox{with }}~ {\rm{B}}_0 =  {\cal{S}}^{-1} \left( {\rm{Q}}_{(1,0)}^{-1} {\rm{Q}}_{(1,1)}\right).
	\end{equation}
	This equation can be used as {\emph{a determining equation}} for the first order symmetries of the system, and provides us with a conservation law
	\begin{equation} \label{eq:conlaw1}
		\left({\cal{T}}-1\right) \ln \det (\tilde{\rm{R}}_1) = \left({\cal{S}}-1 \right) \ln \det ( {\rm{B}}_0).
	\end{equation}
	
	The second condition follows from the $\cal{S}$-independent term and can be written as
	\begin{subequations} \label{eq:deteq1L2}
		\begin{equation}
			{\cal{T}}\left(\tilde{{\rm{R}}}_0 \right)  -  {\rm{B}}_0^{-1} \tilde{{\rm{R}}} _0 {\rm{B}}_0  = {\rm{B}}_0^{-1} \tilde{{\rm{R}}}_1 {\cal{S}}({\rm{B}}_1)   -  {\rm{B}}_0^{-1}  {\rm{B}}_1 {\cal{S}}^{-1}\left( {\rm{B}}_0^{-1} \tilde{{\rm{R}}}_1   {\cal{S}}({\rm{B}}_1) \right) {\rm{B}}_1^{-1} {\rm{B}}_0 ,
		\end{equation}
		where we have used that ${\rm{A}}_1 = -  {\rm{B}}_0^{-1}  {\rm{B}}_1 {\cal{S}}^{-1}\left( {\rm{B}}_0^{-1}\right)$, ${\rm{B}}_0$ is given in \eqref{eq:deteq1L} and 
		\begin{equation}
			{\rm{B}}_1 = {\cal{S}}^{-1}\left( {\rm{Q}}_{(1,0)}^{-1} \left\{ {\rm{Q}}_{(0,1)} - {\rm{Q}}_{(0,0)} {\cal{S}}^{-1} \left(  {\rm{Q}}_{(1,0)}^{-1} {\rm{Q}}_{(1,1)} \right)  \right\} \right).
		\end{equation}
	\end{subequations}
	A conservation law follows from \eqref{eq:deteq1L2} by taking the trace, namely
	\begin{equation}\label{eq:conlaw1a}
		\left({\cal{T}}-1\right) {\rm{Tr}} (\tilde{\rm{R}}_0) = \left({\cal{S}}-1 \right){\cal{S}}^{-1} {\rm{Tr}}\left( {\rm{B}}_0^{-1} \tilde{{\rm{R}}}_1 {\cal{S}}({\rm{B}}_1)\right).
	\end{equation}
	
	On the other hand, using \eqref{eq:defAB} and the Taylor series \eqref{eq:invT}, we can compute ${\cal{A}}^{-1}_T \circ {\cal{B}}$ and ${\cal{B}}^{-1}_T \circ {\cal{A}}$. If we write these operators as
	$${\cal{A}}^{-1}_T \circ {\cal{B}} = {\rm{G}}_0 + {\rm{G}}_1 {\cal{S}} + {\rm{G}}_2 {\cal{S}}^{2} + \ldots, ~~~ {\cal{B}}^{-1}_T \circ {\cal{A}} = {\rm{F}}_0 + {\rm{F}}_1 {\cal{S}} + {\rm{F}}_2 {\cal{S}}^{2} + \ldots, $$
	then equation \eqref{eq:recop2b} becomes
	$${\cal{T}}\left( \hat{\rm{R}}_{-1} \right) {\cal{S}}^{-1} + {\cal{T}}\left(\hat{{\rm{R}}}_0 \right) +   \ldots = \hspace{16cm}$$
	$$ \hspace{2cm} {\rm{G}}_0^{-1}\hat{{\rm{R}}}_{-1} {\cal{S}}^{-1}({\rm{G}}_0) {\cal{S}}^{-1} + \left( {\rm{G}}_0^{-1} \hat{{\rm{R}}}_{-1} {\cal{S}}^{-1}({\rm{G}}_1)  +  {\rm{G}}_0^{-1} \hat{{\rm{R}}} _0 {\rm{G}}_0 + {\rm{F}}_1 {\cal{S}}( \hat{{\rm{R}}}_{-1} ) {\rm{G}}_0 \right) + \ldots,$$
	where ${\rm{F_0G_0 = G_0F_0 = I}}$ was taken into account.  Working as before  we can derive another set of integrability conditions, the first one of which is 
	\begin{equation} \label{eq:deteq1T}
		{{\cal{T}}\left( \hat{\rm{R}}_{-1} \right)}= {\rm{G}}_0^{-1}\hat{{\rm{R}}}_{-1} {\cal{S}}^{-1}({\rm{G}}_0), ~~{\mbox{with }}~ {\rm{G}}_0 =   {\rm{Q}}_{(0,0)}^{-1} {\rm{Q}}_{(0,1)}.
	\end{equation}
	This equation may be used along with \eqref{eq:deteq1L} as a {\emph{determining equation}} for the first order symmetries of the system, and provides a conservation law
	$$ \left({\cal{T}}-1\right) \ln \det (\hat{\rm{R}}_{-1}) = \left({\cal{S}}-1 \right) \ln \det ( {\rm{G}}_0), $$
	which is equivalent to the previous conservation law \eqref{eq:conlaw1}.
	
	The second integrability condition can be written as
	\begin{subequations} \label{eq:deteq1T2}
		\begin{equation}
			{\cal{T}}\left(\hat{{\rm{R}}}_0 \right)  -  {\rm{G}}_0^{-1} \hat{{\rm{R}}} _0 {\rm{G}}_0  = {\rm{G}}_0^{-1} \hat{{\rm{R}}}_{-1} {\cal{S}}^{-1}({\rm{G}}_1)   -  {\rm{G}}_0^{-1}  {\rm{G}}_1 {\cal{S}}\left( {\rm{G}}_0^{-1} \hat{{\rm{R}}}_1   {\cal{S}}^{-1}({\rm{G}}_1) \right) {\rm{G}}_1^{-1} {\rm{G}}_0 ,
		\end{equation}
		where we have used that ${\rm{F}}_1 = -  {\rm{G}}_0^{-1}  {\rm{G}}_1 {\cal{S}}\left( {\rm{G}}_0^{-1}\right)$, ${\rm{G}}_0$ is given in \eqref{eq:deteq1T} and 
		\begin{equation}
			{\rm{G}}_1 =  {\rm{Q}}_{(0,0)}^{-1} \left( {\rm{Q}}_{(1,1)} - {\rm{Q}}_{(1,0)} {\cal{S}} \left(  {\rm{Q}}_{(0,0)}^{-1} {\rm{Q}}_{(0,1)} \right)  \right).
		\end{equation}
	\end{subequations}
	A conservation law follows from \eqref{eq:deteq1T2} by taking the trace, namely
	\begin{equation}\label{eq:conlaw2a}
		\left({\cal{T}}-1\right) {\rm{Tr}} (\hat{\rm{R}}_0) = -\left({\cal{S}}-1 \right) {\rm{Tr}}\left( {\rm{G}}_0^{-1} \hat{{\rm{R}}}_{-1} {\cal{S}}({\rm{G}}_1)\right).
	\end{equation}

	\subsubsection{First example}
	We can apply the above analysis to system \eqref{eq:sys1}, the Jacobian matrices of which are non-singular, see \eqref{eq:matQex1}. We can substitute all these matrices into \eqref{eq:deteq1L}, \eqref{eq:deteq1T} and derive equations for ${\tilde{\rm{R}}}_1$ and ${\hat{\rm{R}}}_{-1}$, respectively. These equations can be solved using the method developed in \cite{X1} by exploiting the eliminable and dynamical variables for system \eqref{eq:sys1}. The relation of ${\tilde{\rm{R}}}_1$, ${\hat{\rm{R}}}_{-1}$ with the symmetries of the system, see \eqref{eq:RLsym} and \eqref{eq:RTsym}, allow us to construct the symmetry up to arbitrary functions of $n$, $u_{0,0}$ and $v_{0,0}$. Finally we specify these functions by employing determining equations \eqref{eq:deteq} which lead to two symmetries, namely
	\begin{equation} \label{eq:sym1}
		\frac{\partial u_{0,0}}{\partial t_1} = \frac{1}{v_{1,0}-v_{-1,0}} , \qquad \frac{\partial v_{0,0}}{\partial t_1} =  \frac{1}{u_{1,0}-u_{-1,0}}\,,
	\end{equation}
	and
	\begin{equation} \label{eq:sym1b}
		\frac{\partial u_{0,0}}{\partial s_1} = \frac{n}{v_{1,0}-v_{-1,0}} + \frac{u_{0,0}}{2 (\alpha- \beta)}, \qquad \frac{\partial v_{0,0}}{\partial s_1} =  \frac{n}{u_{1,0}-u_{-1,0}}+ \frac{v_{0,0}}{2 (\alpha- \beta)}\,,
	\end{equation}
	as well as to the conservation law 
	$$\left(\rho,\sigma\right) =  \left(\ln (u_{1,0}-u_{-1,0}) (v_{1,0}-v_{-1,0}),~  \ln (u_{0,1}-u_{-1,0}) (v_{0,1}-v_{-1,0})\right).$$
	If we consider extended symmetries acting on the parameter $\alpha$, i.e. employ equation \eqref{eq:extdeteq} instead of \eqref{eq:deteq}, we derive the above two symmetries and the extended non-autonomous  symmetry
	\begin{equation}\label{eq:msym1}
		\frac{\partial u_{0,0}}{\partial \tau} = \frac{n}{v_{1,0}-v_{-1,0}}, \qquad \frac{\partial v_{0,0}}{\partial \tau} =  \frac{n}{u_{1,0}-u_{-1,0}}, \qquad 	\frac{\partial \alpha}{\partial \tau} = -1.
	\end{equation}
	The extended symmetry plays the role of a master symmetry for \eqref{eq:sym1} and we can use it to generate an infinite hierarchy of symmetries, and hence prove that system \eqref{eq:sys1} is integrable.

	\subsection{Second case: singular Jacobian matrices} \label{sec:singular}
	
	In this section we consider systems for which not all of the Jacobian matrices ${\rm{Q}}_{(i,j)}$ are invertible, such as systems \eqref{eq:sys2} and \eqref{eq:sys3}. In this case the analysis of equations \eqref{eq:recop2} differs essentially from the previous case as it is more involved depending heavily on the form of the formal series we can use. We demonstrate this situation below by considering two cases which correspond to systems \eqref{eq:sys2} and \eqref{eq:sys3}. These considerations can be easily extended to other systems with singular Jacobians in a similar way. 
	
	We start by considering systems for which ${\rm{Q}}_{(0,0)}$ and ${\rm{Q}}_{(1,1)}$ are singular. In this case we can compute the formal Laurent inverse of operator ${\cal{A}} = {\rm{Q}}_{(0,0)} + {\rm{Q}}_{(1,0)} {\cal{S}}$ using \eqref{eq:AinvL}, and the Taylor inverse of  $ {\cal{B}} =  {\rm{Q}}_{(0,1)} + {\rm{Q}}_{(1,1)} {\cal{S}}$ using \eqref{eq:BinvT}. With only these formal series at our disposal we rearrange the equations in \eqref{eq:recop2} and write them as
	\begin{equation*}
		{\cal{A}}^{-1}_L\circ {\cal{B}}\circ {\cal{T}}({\mathfrak{R}}_L)  = {\mathfrak{R}}_L\circ {\cal{A}}^{-1}_L \circ {\cal{B}} ~~{\mbox{and}}~~  {\cal{T}}({\mathfrak{R}}_T)\circ{\cal{B}}^{-1}_T\circ {\cal{A}}   = {\cal{B}}^{-1}_T\circ {\cal{A}} \circ {\mathfrak{R}}_T,
	\end{equation*}
	respectively. Upon the substitution of the aforementioned series and working as before, we end up with two sets of integrability conditions. The first of these conditions are
	\begin{subequations} \label{eq:deteq2}
		\begin{eqnarray}
			&& {\rm{A}}_0 {\cal{T}}\left(\tilde{{\rm{R}}}_1\right) = \tilde{{\rm{R}}}_1 {\cal{S}}\left({\rm{A}}_0\right),~~ {\mbox{with }} {\rm{A}}_0 = {\cal{S}}^{-1} \left( {\rm{Q}}_{(1,0)}^{-1} {\rm{Q}}_{(1,1)} \right), \label{eq:deteq2a} \\
			&& {\cal{T}}\left(\hat{{\rm{R}}}_{-1}\right) {\cal{S}}^{-1}({\rm{B}}_0 ) = {\rm{B}}_0 \hat{{\rm{R}}}_{-1},~~ {\mbox{with }} {\rm{B}}_0 =  {\rm{Q}}_{(0,1)}^{-1} {\rm{Q}}_{(0,0)}, \label{eq:deteq2b}
		\end{eqnarray}
	\end{subequations}
	which serve also as determining equations for the symmetries of the system under consideration. Moreover, since matrices $\rm{A}_0$ and $\rm{B}_0$ are singular, conservation laws follow from certain entries of determining equations \eqref{eq:deteq2}.  
	
	On the other hand, if all the Jacobian matrices are singular, then ${\cal{A}}^{-1}$ and ${\cal{B}}^{-1}$ might be local and we have to use \eqref{eq:invall} for their computation. In this case we can easily compute the compositions
	\begin{equation}\label{eq:compAB}
		{\cal{B}}^{-1}\circ {\cal{A}} = {\rm{A}}_{-1} {\cal{S}}^{-1} + {\rm{A}}_{0} + {\rm{A}}_{1} {\cal{S}},\qquad {\cal{A}}^{-1}\circ {\cal{B}} = {\rm{B}}_{-1} {\cal{S}}^{-1} + {\rm{B}}_{0} + {\rm{B}}_{1} {\cal{S}}.
	\end{equation}
	Substituting them into \eqref{eq:recop2a} and working as before, we can derive an infinite set of conditions, the first three of which are the following ones.
	\begin{subequations}\label{eq:detallL}
		\begin{eqnarray}
			&& {\rm{A}}_{1}{\cal{S}}(\tilde{{\rm{R}}}_1){\cal{S}}^{2} ( {\rm{B}}_{1}) = {\rm{O}}, \label{eq:detallL1}\\ 
			&& {\rm{A}}_{1}{\cal{S}}(\tilde{{\rm{R}}}_1){\cal{S}}^{2} ( {\rm{B}}_{0} ) +  ({\rm{A}}_{0} \tilde{{\rm{R}}}_1 + {\rm{A}}_{1} {\cal{S}}( \tilde{{\rm{R}}} _0)){\cal{S}}({\rm{B}}_{1}) = {\rm{O}}, \label{eq:detallL2}\\
			&&{\cal{T}}(\tilde{{\rm{R}}}_1) =  {\rm{A}}_{1} {\cal{S}}( \tilde{{\rm{R}}} _1) {\cal{S}}^2({\rm{B}}_{-1}) + ({\rm{A}}_{0} \tilde{{\rm{R}}}_1 + {\rm{A}}_{1} {\cal{S}}( \tilde{{\rm{R}}} _0)){\cal{S}}({\rm{B}}_{0})   \nonumber \\
			&& \hspace{2.3cm} +  ( {\rm{A}}_{-1} {\cal{S}}^{-1}(\tilde{{\rm{R}}}_1)+ {\rm{A}}_{0}\tilde{{\rm{R}}} _0  + {\rm{A}}_{1} {\cal{S}}(\tilde{{\rm{R}}}_{-1})) {\rm{B}}_{1}. \label{eq:detallL3}
		\end{eqnarray}
	\end{subequations}
	Similarly, substituting \eqref{eq:compAB} into \eqref{eq:recop2b} we end up with another infinite set of conditions, the first three of which are
	\begin{subequations} \label{eq:detallT}
		\begin{eqnarray}
			&&{\rm{A}}_{-1}{\cal{S}}^{-1}(\hat{{\rm{R}}}_{-1}){\cal{S}}^{-2} ( {\rm{B}}_{-1}) = {\rm{O}}, \label{eq:detallT1}\\ 
			&&{\rm{A}}_{-1}{\cal{S}}^{-1}(\hat{{\rm{R}}}_{-1}){\cal{S}}^{-2} ( {\rm{B}}_{0} ) +  ({\rm{A}}_{0} \hat{{\rm{R}}}_{-1} + {\rm{A}}_{-1} {\cal{S}}^{-1}( \hat{{\rm{R}}} _0)){\cal{S}}^{-1}({\rm{B}}_{-1}) = {\rm{O}}, \label{eq:detallT2}\\
			&& {\cal{T}}(\hat{{\rm{R}}}_{-1}) =  {\rm{A}}_{-1} {\cal{S}}^{-1}( \hat{{\rm{R}}} _{-1}) {\cal{S}}^{-2}({\rm{B}}_{1})  + ({\rm{A}}_{0} \hat{{\rm{R}}}_{-1} + {\rm{A}}_{-1} {\cal{S}}^{-1}( \hat{{\rm{R}}} _0)){\cal{S}}^{-1}({\rm{B}}_{0})\nonumber \\
			&&\hspace{2.3cm} + ({\rm{A}}_{-1} {\cal{S}}^{-1}(\hat{{\rm{R}}}_1) + {\rm{A}}_{0}\hat{{\rm{R}}} _0  + {\rm{A}}_{1} {\cal{S}}(\hat{{\rm{R}}}_{-1})) {\rm{B}}_{-1}. \label{eq:detallT3}
		\end{eqnarray}
	\end{subequations}
	Matrices ${\rm{A}}_i$ and ${\rm{B}}_i$ involved in both integrability conditions \eqref{eq:detallL} and \eqref{eq:detallT} are the ones appearing in \eqref{eq:compAB}. Moreover, all the matrices in these integrability conditions are singular. Equations \eqref{eq:detallL1} and \eqref{eq:detallT1} partially determine the leading terms in the formal series of the recursion operator, and conservation laws may follow from certain entries of the equations \eqref{eq:detallL3} and \eqref{eq:detallT3}.

	\subsubsection{Second example}
	
	Matrices $ {\rm{Q}}_{(0,0)}$ and $ {\rm{Q}}_{(1,1)}$ for system \eqref{eq:sys2} are singular, see \eqref{eq:matQex2}, so we can use equations  \eqref{eq:deteq2}. Working as in the previous example, we substitute matrices \eqref{eq:matQex2} into \eqref{eq:deteq2} to derive equations for the leading terms of the Laurent and Taylor series. On one hand, the analysis of these equations yield the conservation law 
	$(\rho,\sigma) =  \left( \ln (1+ u_{-1,0} v_{1,0}),~  \ln (1+ u_{-1,0} v_{0,1})\right)$. On the other hand, their solutions determine the symmetries up to some arbitrary functions of $n$, $u_{0,0}$ and $v_{0,0}$. The substitutions of these solutions into \eqref{eq:extdeteq} lead to the symmetry
	\begin{equation} \label{eq:sym2}
		\frac{\partial u_{0,0}}{\partial t_1} =  \frac{u_{-1,0}}{1+ u_{-1,0} v_{1,0}} , \qquad \frac{\partial v_{0,0}}{\partial t_1}  = \frac{-v_{1,0}}{1+ u_{-1,0} v_{1,0}},
	\end{equation}
	and the extended symmetry
	\begin{equation}\label{eq:msysm2}
		\frac{\partial u_{0,0}}{\partial \tau}  =  \frac{n\,u_{-1,0}}{1+ u_{-1,0} v_{1,0}}, \qquad \frac{\partial v_{0,0}}{\partial \tau}  =  \frac{-n\,v_{1,0}}{1+ u_{-1,0} v_{1,0}}, \qquad 	\frac{\partial \alpha}{\partial \tau} = 1,
	\end{equation}
	which plays the role of a master symmetry \eqref{eq:sym2}, and thus proves the integrability of system \eqref{eq:sys2}.

	\subsubsection{Third example} \label{sec:3exsym}
	
	For system \eqref{eq:sys3} all the Jacobian matrices are singular, thus ${\cal{A}}^{-1}$ and  ${\cal{B}}^{-1}$ are local, i.e.
	$${\cal{A}}^{-1} =  \begin{pmatrix}
		0 & 0 \\
		\tfrac{f_{-1,0}}{u_{-1,0}-u_{0,0}} & \tfrac{1}{u_{0,0}-u_{-1,0}}
	\end{pmatrix} {\cal{S}}^{-1} +  \begin{pmatrix} 1 & 0 \\ \tfrac{v_{0,0}-u_{-1,0}}{u_{-1,0}-u_{0,0}} & 0 \end{pmatrix},~~ {\cal{B}}^{-1} =  \begin{pmatrix} \tfrac{u_{0,0}-u_{0,1}}{u_{0,0}-v_{0,1}} & 0 \\ -1 & 0 \end{pmatrix} {\cal{S}}^{-1} + \begin{pmatrix}
		0 & \tfrac{1}{u_{0,0}-v_{0,1}}  \\ 0 & 0
	\end{pmatrix},$$
	where $ f_{0,0} = u_{0,1}+v_{0,1} - u_{1,0}-v_{1,0}$. We can easily compute 
	$${\cal{B}}^{-1}\circ{\cal{A}}  = {\rm{A}}_{-1} {\cal{S}}^{-1} + {\rm{A}}_{0} + {\rm{A}}_{1} {\cal{S}}=  \begin{pmatrix} \tfrac{u_{0,0}-u_{0,1}}{u_{0,0}-v_{0,1}} & 0 \\ -1 & 0 \end{pmatrix} {\cal{S}}^{-1} + \begin{pmatrix} \tfrac{f_{0,0}}{u_{0,0}-v_{0,1}} & 0  \\ 0 & 0 \end{pmatrix} +  \begin{pmatrix} \tfrac{v_{1,0}-u_{0,0}}{u_{0,0}-v_{0,1}} & \tfrac{u_{1,0}-u_{0,0}}{u_{0,0}-v_{0,1}} \\ 0 & 0 \end{pmatrix} {\cal{S}}$$
	and
	$${\cal{A}}^{-1}\circ{\cal{B}}  = {\rm{B}}_{-1} {\cal{S}}^{-1} + {\rm{B}}_{0} + {\rm{B}}_{1} {\cal{S}}=  \begin{pmatrix}0& 0 \\  \tfrac{v_{-1,1}-u_{-1,0}}{u_{-1,0}-u_{0,0}}  & \tfrac{u_{-1,1}-u_{-1,0}}{u_{-1,0}-u_{0,0}}  \end{pmatrix} {\cal{S}}^{-1} + \begin{pmatrix} 0& 0  \\ 0 & \tfrac{f_{-1,0}}{u_{0,0}-u_{-1,0}}  \end{pmatrix} +  \begin{pmatrix}0  & -1 \\ 0 & \tfrac{u_{-1,0}-v_{0,0}}{u_{-1,0}-u_{0,0}} \end{pmatrix} {\cal{S}}.$$
	Taking into account the above relations we employ \eqref{eq:detallL}, \eqref{eq:detallT} along with \eqref{eq:deteq} to find the symmetries of \eqref{eq:sys3}. Moreover by exploiting the entries of the matrix equation \eqref{eq:detallL3} conservation laws can also be derived. Specifically, our considerations yield two generalised symmetries and corresponding conservation laws. The first symmetry is 
	\begin{equation} \label{eq:sym3}
		\frac{\partial u_{0,0}}{\partial t_1} =  \frac{u_{0,0}-v_{1,0}}{u_{-1,0}-v_{1,0}} , \qquad \frac{\partial v_{0,0}}{\partial t_1}  =   \frac{u_{-1,0}-v_{0,0}}{u_{-1,0}-v_{1,0}},
	\end{equation}
	and the corresponding conservation law is
	$$ \left(\rho_1,~\sigma_1 \right)=  \left( \ln \frac{v_{0,0}-u_{-1,0}}{(v_{1,0}-u_{-1,0})^2},~~  \ln \frac{(u_{-2,0}-v_{0,0})^2}{u_{-2,0}-u_{-1,0}}\right).$$
	The second symmetry is
	\begin{equation}\label{eq:sym3a}
		\frac{\partial u_{0,0}}{\partial s_1}  =  (u_{1,0}-u_{0,0})(v_{1,0}-u_{0,0}), \qquad \frac{\partial v_{0,0}}{\partial s_1} = (u_{-1,0}-v_{0,0}) (v_{-1,0}-v_{0,0}),
	\end{equation}
	and the corresponding conservation law is $ \left(\varrho,\varsigma\right) = \left(\ln (v_{1,0}-u_{0,0}), \ln (v_{0,1}-u_{0,0})\right)$. 
	
	The existence of only two first order symmetries is not sufficient to prove the integrability of system \eqref{eq:sys3}. For this purpose we work as proposed in Remark \ref{rem:Rell} and try to find higher order symmetries. Starting with symmetry \eqref{eq:sym3} along with \eqref{eq:RLsym} and \eqref{eq:RTsym}, we consider equations \eqref{eq:recop2} for ${\mathfrak{R}}^\ell_L$ and ${\mathfrak{R}}^\ell_T$, with $\ell=2,3$, and derive determining equations for second and third order symmetries, respectively. In this way we find the second order symmetry
	\begin{subequations} \label{eq:sym3b}
		\begin{eqnarray}
			\frac{\partial u_{0,0}}{\partial t_2} &=&  \frac{u_{0,0}-v_{1,0}}{(u_{-1,0}-v_{1,0})^2} \left( \frac{u_{0,0}-u_{-1,0}}{v_{2,0}-u_{0,0}} + \frac{u_{-1,0}-v_{0,0}}{v_{0,0}-u_{-2,0}} + 1\right), \\ 
			\frac{\partial v_{0,0}}{\partial t_2}  &=& \frac{u_{-1,0}-v_{0,0}}{(u_{-1,0}-v_{1,0})^2}  \left( \frac{u_{0,0}-v_{1,0}}{v_{2,0}-u_{0,0}} + \frac{v_{1,0}-v_{0,0}}{v_{0,0}-u_{-2,0}} + 1\right),
		\end{eqnarray}
	\end{subequations}
	and a third order one which is omitted here because of its length. Both these symmetries commute with \eqref{eq:sym3} and with each other. Another product of our considerations is the conservation law 
	$$ \left(\rho_2,\sigma_2 \right)=  \left( \frac{v_{1,0}-v_{0,0}+u_{-1,0}-u_{-2,0}}{(v_{0,0}-u_{-2,0}) (v_{1,0}-u_{-1,0})},  \frac{1}{v_{-1,0}-u_{-3,0}}\left( \frac{u_{-2,0}-u_{-3,0}}{u_{-1,0}-u_{-2,0}} + \frac{v_{-1,0} -2 u_{-2,0} + u_{-3,0}}{v_{0,0}-u_{-2,0}} \right) \right).$$
	derived from the integrability condition for $\tilde{{\rm{R}}}_0$. 
	
	We can use the same approach starting with \eqref{eq:sym3a} to find the second order symmetry
	\begin{subequations} \label{eq:sym3c}
		\begin{eqnarray}
			\frac{\partial u_{0,0}}{\partial s_2} &=&  (u_{0,0}-v_{1,0}) \left( u_{2,0} v_{2,0} - (u_{2,0} + v_{2,0} - v_{0,0}) u_{1,0} + (u_{1,0}-v_{0,0}) u_{0,0}\right), \\ 
			\frac{\partial v_{0,0}}{\partial s_2}  &=& (u_{-1,0}-v_{0,0})\left( u_{-2,0} v_{-2,0} - (u_{-2,0}+ v_{-2,0} - u_{0,0}) v_{-1,0} + (v_{-1,0}-u_{0,0}) v_{0,0}\right),
		\end{eqnarray}
	\end{subequations}
	and a third order one which again is omitted because of its length. Both these symmetries commute with \eqref{eq:sym3} and with each other. In fact all six symmetries we found commute with each other. The existence of these two sets of commuting symmetries is a strong indication that system \eqref{eq:sys3} is integrable.
	
	\begin{remark}
		All our above considerations and results on symmetries in the $n$ direction, $\cal{S}$-pseudo-difference operators and corresponding integrability conditions can be easily modified in order to discuss first order symmetries in the $m$ direction, $\cal{T}$-pseudo-difference operators, and derive corresponding integrability conditions. Indeed it is sufficient to change $ {\bs{u}}_{i,j}$ to ${\bs{u}}_{j,i}$,  ${\rm{Q}}_{(i,j)}$ to ${\rm{Q}}_{(j,i)}$, and to interchange variables $n$ and $m$, as well as the shift operators ${\cal{S}}$ and ${\cal{T}}$ everywhere in recursion operators, formal series and integrability conditions we presented in this section. In particular, for the three systems \eqref{eq:sys1}, \eqref{eq:sys2} and \eqref{eq:sys3} we considered here, it can be easily verified that they are invariant under the interchanges $(u_{i,j},v_{i,j}) \longleftrightarrow (u_{j,i},v_{j,i})$ and $(\alpha,\beta) \longleftrightarrow (\beta,\alpha)$. A consequence of this invariance is that symmetries in the second lattice direction follow from the symmetries we presented here by applying the aforementioned interchanges.
	\end{remark}
	
	\section{Conclusions}
	
	In this paper we considered quad systems which satisfy the requirements of Definition \ref{def:class} and discussed the elimination of variables using the quad system and its shifts as well as its relation to the initial value problem. We gave a short introduction to recursion operators and integrability conditions and presented a systematic method for computing the symmetries of quad systems which exploits formal recursion operators and integrability conditions. Specifically, we analysed equation \eqref{eq:recop1} and derived necessary integrability conditions which can be interpreted as determining equations for symmetries. We considered three particular systems from the class $\cal{Q}$ under consideration which highlight the differences and the similarities between the multicomponent and the scalar case, but also the differences among different members of the class of quad systems $\cal{Q}$.
	
	It is evident that there is a close connection between symmetries and integrability however, it would be interesting to develop a method for the derivation of generalised symmetries without the use of the theory of integrability conditions. This would require the exploitation of all possible choices of eliminable variables for the analysis of equations \eqref{eq:deteq} and \eqref{eq:extdeteq} which would result in the derivation of additional determining equations, similar to the integrability conditions we presented here. The advantage of this approach would be that it will not require the machinery of pseudo-difference operators and formal series, and it will be easier to implement in symbolic computations. Moreover having computed the lowest order symmetries in this way, it would be easier then to prove the integrability of a quad system by using the machinery we presented in Section \ref{sec:recop}.
	
	Finally, it would be interesting to employ generalised symmetries in the construction of solutions, such as group invariant solutions. In particular, the non-autonomous symmetries, like \eqref{eq:msym1} and \eqref{eq:msysm2}, are expected to lead to non-autonomous integrable maps, whereas the non-autonomous ones could lead to discrete Painlev{\'e} type systems. 
	
	\section*{Acknowledgements}
	
	The work of Louis Brady was supported by a London Mathematical Society {\emph{Undergraduate Research Bursary}} grant (Grant No URB--2021--49).

\end{document}